# The advancement of Brillouin Light Scattering with the assistance of nanoplasmonic structures. Enhancement and amplification

Eugene Bortchagovsky[1], Andrii V. Chumak[2]* and Valeri Lozovski[3]**

[1] V. Lashkaryov Institute of Semiconductor Physics of NAS of Ukraine, 41 Nauki Avenue, Kyiv 03028, Ukraine
[2] Faculty of Physics, University of Vienna, Boltzmanngasse 5 A, 1090 Vienna, Austria
[3] Educational Scientific Institute of High Technologies, Taras Shevchenko National University of Kyiv, 4-g Hlushkova Avenue, Kyiv, 03022, Ukraine

**Abstract**

Brillouin light scattering (BLS) is a key technique in studying magnonic systems, but its sensitivity is often limited. While nanoplasmonic systems can enhance BLS through near-field effects, we propose a novel approach for additional amplification. In this conceptual paper, we show how to actively supply energy to a surface collective electromagnetic resonance (SCR) supported by a sparse layer of metal nanoparticles on a magnetic film. Proposed methods are designed to significantly amplify the efficiency of surface-enhanced Brillouin light scattering without increasing the intensity of the primary excitation. In the proposed scheme, the pump extends the propagation length of the SCR, leading directly to BLS amplification. We analyze the conditions for such amplification, with numerical estimates indicating a potential gain of more than an order of magnitude in the surface-wave amplitude. This gain far surpasses the modest increase achievable through passive enhancement alone. These findings outline a practical pathway to achieving BLS amplification in integrated magnonic platforms. The feasibility of the proposed approaches is discussed in light of recent experimental and theoretical advances in plasmonics, magneto-plasmonics, and magnonics.

Corresponding authors e-mails:

*A.V. Chumak: andrii.chumak@univie.ac.at
**V. Lozovski: vlozovski@knu.ua

## I. INTRODUCTION

Magnonics has emerged as a promising branch of magnetism [1–4], with notable progress toward energy-efficient data processing. For example, ref. [5] reports a device integrating various radio-frequency components that operates in the gigahertz range. Research in magnonics requires control of spin-wave (magnon) properties, and Brillouin light scattering (BLS) is an effective tool for this purpose [1]. The advantages of BLS for studying magnetic films include: (a) it is non-destructive; (b) it offers high sensitivity; (c) it covers a broad frequency range of magnetic excitations, enabling studies of different magnons and their dynamics [6] – BLS is widely used for soft magnetic alloys such as Fe-Ni [7] and for yttrium iron garnet (YIG) [8–10]; (d) it provides high frequency resolution, allowing precise determination of magnetic excitation energies; owing to micro-BLS, magnetic properties can be probed at the submicron scale with spatial maps of spin-wave intensity; (e) it enables direct determination of magnon wave vectors, essential for dispersion studies; (f) it directly probes spin waves—their frequencies, intensities, damping and transport – in both linear and nonlinear regimes; and (g) it can be combined with complementary methods such as the magneto-optical Kerr effect (MOKE), ferromagnetic resonance (FMR) and scanning near-field optical microscopy (SNOM). Overall, BLS is a powerful and versatile technique for investigating the static and dynamic magnetic properties of thin films across multiple spatial and temporal scales, providing crucial input for spintronics and magnetic nanotechnologies.

Despite its advantages, BLS spectroscopy suffers from inherently weak signals, which limits its broader use. Enhancing the signal to experimentally and practically acceptable levels is therefore essential. Several approaches have been explored, including plasmonic structures on magnetic films [11–13], thin-film optical waveguides [14,15], photonic crystals [16,17] and Mie-resonators [18,19].

Previous studies [12,13] considered BLS enhancement arising from photon–magnon interaction strengthened by a nanoplasmonic system on the film surface. The derived analytical expressions show that resonant oscillations in the nanoplasmonic system increase the BLS intensity. This electromagnetic mechanism is equivalent to the enhancement observed in surface-enhanced Raman scattering (SERS) [20,21]. Resonant excitation of the plasmonic system is a prerequisite; optimally, both the incident and scattered fields are enhanced. In Raman scattering the two frequencies can be widely separated, often requiring a broad resonance and reducing efficiency. By contrast, BLS involves a small frequency shift, which makes resonance-based plasmonic enhancement more favorable.

The resonance conditions for a layer of interacting nanoparticles differ from the localised plasmon resonance (LPR) of an isolated particle due to electromagnetic coupling [22,23]. To avoid confusion with surface plasmon resonance (SPR) at a material interface [24], we refer to the collective resonance of a layer of randomly distributed nanoparticles as a surface common resonance (SCR), in analogy with the surface lattice resonance (SLR) of ordered arrays [11,25]. Starting from the dispersionless LPR of well-separated particles, increasing surface coverage modifies the dispersion of SCR modes, which approach the electromagnetic modes of a continuous metallic layer in an asymmetric environment. A key distinction from SPR is that SCR can be excited directly by external radiation without additional wave-vector matching, whereas SPR typically cannot. However, as well as LPR, SCR has its dispersion as in the range of evanescent fields $k > \omega/c$ [26,27], as in the range of propagating light $k < \omega/c$ [28,29]. It allows to use different ways of SCR excitation.

Resonance conditions correspond to an increase in the effective susceptibility of the system [12,13]. This susceptibility is often written in the form $A/B$, for example in a Lorentz-type response $P/((\omega-\omega_0)^2-i\gamma\omega)$, where $P$ is the resonance strength and $\omega_0$ and $\gamma$ are the resonance frequency and damping, respectively. The resonance condition is $\mathrm{Re}(B)=0$, so at resonance the susceptibility becomes $A/\mathrm{Im}(B)$, indicating two routes to enhancement (or a higher quality factor). The common approach is to reduce damping by introducing additional interactions that partially compensate intrinsic losses, as in operation at Rayleigh anomalies for ordered lattices of plasmonic nanoparticles [11,25]. An alternative is to increase the resonance strength by raising the numerator $A$ through pumping that supplies additional energy from other processes within the system.

Thus, the amplitude of the BLS signal follows the value of the effective susceptibility of a nanoplasmonic system deposited on a surface of a magnetic film and increases by tuning the system into plasmonic resonance. We want to distinguish two different mechanisms for BLS signal increasing. Enhancement – the natural approach utilizes the local field enhancement happened at plasmonic system excited by an external illumination as discussed in refs. [12,13]. Amplification

– the approach, which uses some additional channel supplying additional external energy to the plasmonic system to amplify its susceptibility, thereby amplifying the natural enhancement and the BLS signal correspondingly.

In this work, we consider nanoplasmonic structures supporting collective electromagnetic modes. Our focus is on enhancing such modes by transferring energy from an external source. The problem is naturally addressed within the instability theory for distributed-parameter systems and requires only the wave dispersion equation [30] (see Supplementary information). According to this theory, two instability types can arise: absolute and convective. A system with absolute instability can act as a generator, whereas a system with convective instability functions as an amplifier [31–33]. As is well known, amplification requires external energy input. Many studies of intrinsic-excitation amplification in solids have used a direct current (DC) as the energy source [34–36]. In particular, Ref. [36] examines localized modes in a three-layer semiconductor structure carrying current; dispersion analysis shows that, at a certain carrier velocity, energy is transferred from longitudinal space-charge waves – powered by the external source – to film modes of plasmon-polariton character, formally demonstrating convective instability. However, DC-based BLS amplification can be technologically impractical. In such cases, energy can instead be supplied by transferring energy to SCR via appropriate auxiliary absorbing structures on the magnetic-film surface excited by an additional light source.

This work analyses the feasibility of this approach via energy transfer either from a current flowing in the substrate or from additional absorbers—such as semiconductor quantum dots or organic dye molecules – co-deposited with plasmonic nanoparticles on the surface. The feasibility of the proposed amplification mechanisms is discussed in Section V in light of experimental and theoretical results from related fields.

## II. BASIS FORMALISM FOR BLS ENHANCEMENT BY A NANOPLASMONIC SYSTEM

A model for dynamic magnetisation excitation in a magnetic medium—producing a small, frequency-dependent perturbation of the susceptibility tensor – was presented in [37]. For scattering from magnons (spin waves), this perturbation arises from magneto-optical coupling. The magnetisation-dependent part of the polarisation $P_i$ induced by the incident electric field can be written as:

$$P_i(\mathbf{R},\omega\pm\Omega) = \left(1/4\pi\right)\varepsilon_0\chi_{ij}(\mathbf{R},\Omega)E_j(\mathbf{R},\omega) \ , \tag{1}$$

where $E_j(\mathbf{R},\omega)$ is a component of the electric field inside the film, $\chi_{ij}(\mathbf{R},\Omega)$ is the susceptibility tensor in the first-order magneto-optical effects; its relation to the Kerr effect is given by [37].

When the nanoparticles are located on the magnetic film, the field scattered by the magnons (see, Eq.(1)) will consist of two parts – the initial field $E_i^{(0)}(\mathbf{R},\omega)$, which is inside the magnetic film without a nanoplasmonic cover, and the field reradiated by the nanoplasmonic system $E_i^{(rP)}(\mathbf{R},\omega)$

$$E_i(\mathbf{R},\omega) = E_i^{(0)}(\mathbf{R},\omega) + E_i^{(rP)}(\mathbf{R},\omega), \tag{2}$$

where

$$E_i^{(rP)}(\mathbf{R},\omega) = -k_0^2\sum_{n=1}^{N}\int_V d\mathbf{R}' G_{ij}^{(23)}(\mathbf{R},\mathbf{R}',\omega)\mathrm{X}_{jl}^{(P)}(\mathbf{R}',\omega)E_i^{(ext)}(\mathbf{R},\omega), \tag{3}$$

with $N$ the number of nanoparticles on the film surface, $G_{ij}^{(23)}(\mathbf{R},\mathbf{R}',\omega)$ electrodynamic Green's function of three-layer system [38] – a substrate (1), magnetic film (2), and environment (3) as it is shown in Fig.1. The superscript ($\alpha\beta$) in the Green's function means that the source of the field is in the medium $\beta$=3 (environment) and the observation point is located in the medium $\alpha$=2 (inside the film). $\mathrm{X}_{jl}^{(P)}(\mathbf{R}',\omega)$ in Eq.(3) is an effective susceptibility of the nanoplasmonic cover [39], integration is over the volume of a nanoparticle.

The oscillating electric dipole described by Eq. (1) can be considered as the source of the scattered field at the shifted frequency (i.e., the field of BLS). Further, we consider Stokes shifted BLS at the frequency $\bar{\omega} = \omega - \Omega$. The scattered field at the shifted frequency $\bar{\omega}$ consists of two parts – the field directly scattered by the film

$$E_i^{(BLS-I)}(\mathbf{R},\varpi) = -\bar{k}_0^2 \int_{V_{\text{int}}} d\mathbf{R}' G_{ij}^{(32)}(\mathbf{R},\mathbf{R}',\varpi) P_j(\mathbf{R}',\varpi), \quad (4)$$

where the dipole momentum is calculated by Eq.(1) with the field from Eq.(2). The second part of the scattered field is caused by the scattering by the nanoplasmonic system on the surface

$$E_i^{(BLS-II)}(\mathbf{R},\varpi) = -\bar{k}_0^2 \int_{V_{\text{int}}} d\mathbf{R}' G_{ij}^{(33)}(\mathbf{R},\mathbf{R}',\varpi) \mathrm{X}_{jl}^{(P)}(\mathbf{R}',\varpi) E_l^{(BLS-I)}(\mathbf{R}',\varpi). \quad (5)$$

The enhancement of BLS can be achieved when the effective susceptibilities of the nanoplasmonic cover $\mathrm{X}_{jl}^{(P)}(\mathbf{R}',\omega)$ reach their maximum.

## III. AMPLIFICATION OF BLS BY DC

We first formulate the problem of BLS amplification via energy transfer from a direct current (DC) to SCR. The system considered is a solid substrate, e.g., a semiconductor like InSb, bearing a thin magnetic film, for example yttrium iron garnet (YIG) or permalloy. For the alternative route, where energy is transferred from additional absorbers, we consider conventional stacks with a YIG film on a GGG substrate.

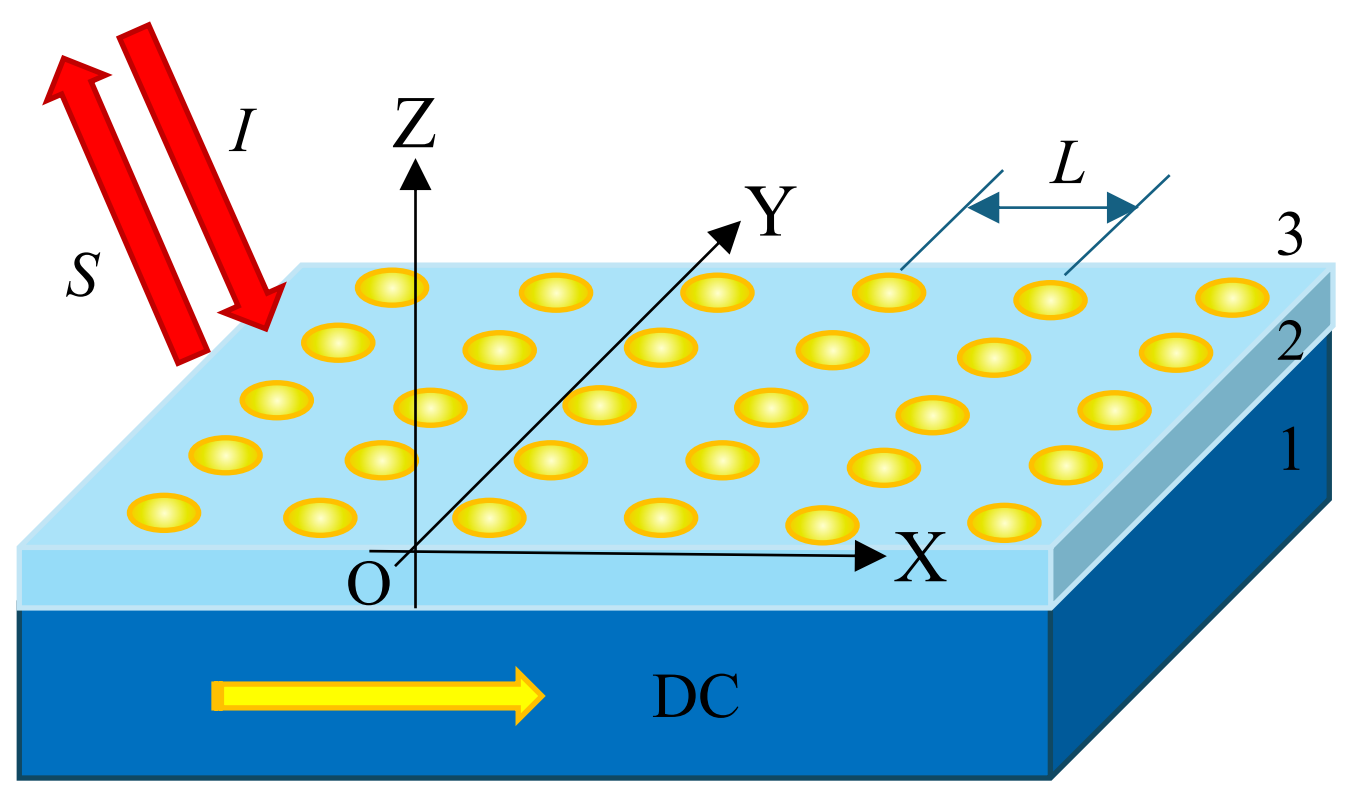

Fig. 1. Setup of the system where the amplification of BLS by a DC in the substrate can be observed.

As noted above and reported in [12,13], it is necessary to create conditions where the effective susceptibility of the submonolayer of nanoparticles reaches a maximum to enhance Brillouin light scattering (BLS). This maximum will correspond to the minimum of the pole part of the effective susceptibility. In this section, we will consider additional possibilities for enhancing BLS by transferring energy from a DC in the substrate to the plasmonic system on a surface of magnetic film. Since the system under consideration is macroscopically homogeneous in the film plane, we can perform a 2D Fourier transform in the film plane and proceed to the so-called **k**-*z* representation. Thus, we will further analyze the effective susceptibility in the **k**-*z* representation $\mathrm{X}_{jl}^{(P)}(\mathbf{k},z,\omega)$, with **k** lying in the plane of the film. The minimum of the pole part $B(\mathbf{k},z,\omega)$ of the effective susceptibility is achieved when $\mathrm{Re}\,B(\mathbf{k},z,\omega)=0$. This condition is nothing more than the dispersion relation of the SCR considered in this work.

For the nanoplasmonic system, which is the layer of nanoparticles randomly and homogeneously distributed along the surface, mathematically, the condition of collective plasmonic resonance can be written as

$$\mathrm{Re}\det\left[\left(\tilde{\chi}_{ij}(\omega)\right)^{-1} + n k_0^2 G_{ji}^{(33)}(\mathbf{k},z_p,z_p,\omega)\right] = 0, \quad (6)$$

where $n$ is the concentration of nanoparticles on the surface, and $\tilde{\chi}_{ij}(\omega)$ is the susceptibility of a single nanoparticle on the surface of the film. Eq.(6) is the dispersion relation for the SCR. When nanoparticles form a regular structure on the surface (2D photonic crystal), the effective susceptibility of the SLR will depend on the reciprocal lattice vector.

Analogously to [36], the convection instability, manifesting the possibility of using the system as an amplifier, can be realized when the phase synchronism domain is located at rather large values of wave vector. Thus, it is necessary to make additional efforts to shift the domain of phase synchronism from large values of wave vectors to smaller ones. This problem can be solved using a planar photonic crystal. Due to the photonic crystal structure on the surface of the magnetic film, the *umklapp* processes lead to the propagation of a surface wave with the shifted wave vector $\mathbf{k}_{sh} = \mathbf{k} - \mathbf{w}$, with $\mathbf{w}$ – the vector of reciprocal lattice along the OX axis of the photonic crystal (minimal value of this vector $\mathbf{w} = (2\pi / L)\mathbf{e}_x$).

Taking these circumstances into account, one can propose a three-layer structure (semiconductor substrate - magnetic film at the surface of which the nanoplasmonic photonic crystal is located - environment), which is illuminated by a testing light beam (Fig. 1). The DC current flows along the semiconductor substrate. The arrows labelled I and S represent the incident and backscattered light, respectively. The conditions for transferring energy from a DC (which is provided by an external energy source) to the surface plasmon imply that the system under consideration is under convective instability. This energy transfer will be the reason for the amplification of BLS, increasing the strength of SLR.

The problem of the possibility of using the system as an amplifier can be solved within the framework of Sturrock's rule [32], which is based on the consideration of the dispersion relations of the waves. Thus, one should obtain the dispersion relations for the SLR that can be excited in the system shown in Fig. 1 Before deriving the dispersion relation, it is necessary to obtain the effective dielectric permittivity of the medium with an electrical current. This problem was solved within the framework of the hydrodynamic approximation for the flow of charge carriers in a semiconductor [36]. One assumes the uniform current distribution, so the substrate changes its permittivity uniformly too. As a result, one obtained

$$\varepsilon_V = \begin{pmatrix} \varepsilon_{xx} & 0 & 0 \\ 0 & \varepsilon_{yy} & 0 \\ 0 & 0 & \varepsilon_{zz} \end{pmatrix}, \tag{7}$$

with

$$\begin{gathered} \varepsilon_{xx} = \frac{\left(\varepsilon_L - \omega_p^2 / \left(W \cdot W'\right)\right)\left(\varepsilon_L - \omega_p^2 / \left(\omega W'\right)\right)}{\varepsilon_L - \omega_p^2\left(1 - \varepsilon_L (V/c)^2\right) / W \cdot W'}, \\ \varepsilon_{yy} = \varepsilon_L - \frac{\omega_p^2 W}{\omega^2 W'}, \qquad \varepsilon_{zz} = \frac{\varepsilon_L - \omega_p^2 / \omega W'}{1 - \omega_p^2 V / k c^2 W'}, \end{gathered} \tag{8}$$

where $W = \omega - kV$, $W' = \omega - kV + i\gamma$, $\omega_p^2 = 4\pi\rho_0 e / m^*$ is the plasma frequency of a semiconductor, $\rho_0$ is the steady-state density of free carriers, and $V$ is their steady-state drift velocity, $\varepsilon_L$ is the semiconductor permittivity, and $\gamma$ is the decay constant of the electronic plasma in a substrate. DC propagates along the OX direction (Fig.1). Now using the dielectric function of the substrate with constant flowing current (7,8) in the Green function of the system we obtain modified equation of type (6), which solution in the case of uniformly arranged nanoparticles on the surface, are shown in Fig. 2. In this scenario, the phase synchronism region (highlighted in yellow), which characterizes the instability region in the system, lies within the range of large wave vectors, corresponding to a wavelength $\lambda \approx 225$ nm. To shift the phase synchronism region into the range of shorter wave vectors and more appropriate wavelengths, a system of nanoparticles regularly arranged on the film's surface, forming a 2D photonic crystal, can be utilized. In this case, to calculate the effective susceptibility, it is necessary to consider the contribution of higher spatial harmonics, i.e., to account for the field at the fundamental spatial harmonic $\mathbf{k}$ and at wave vectors shifted by the reciprocal lattice vector $\mathbf{k} \pm \mathbf{w}$ [36].

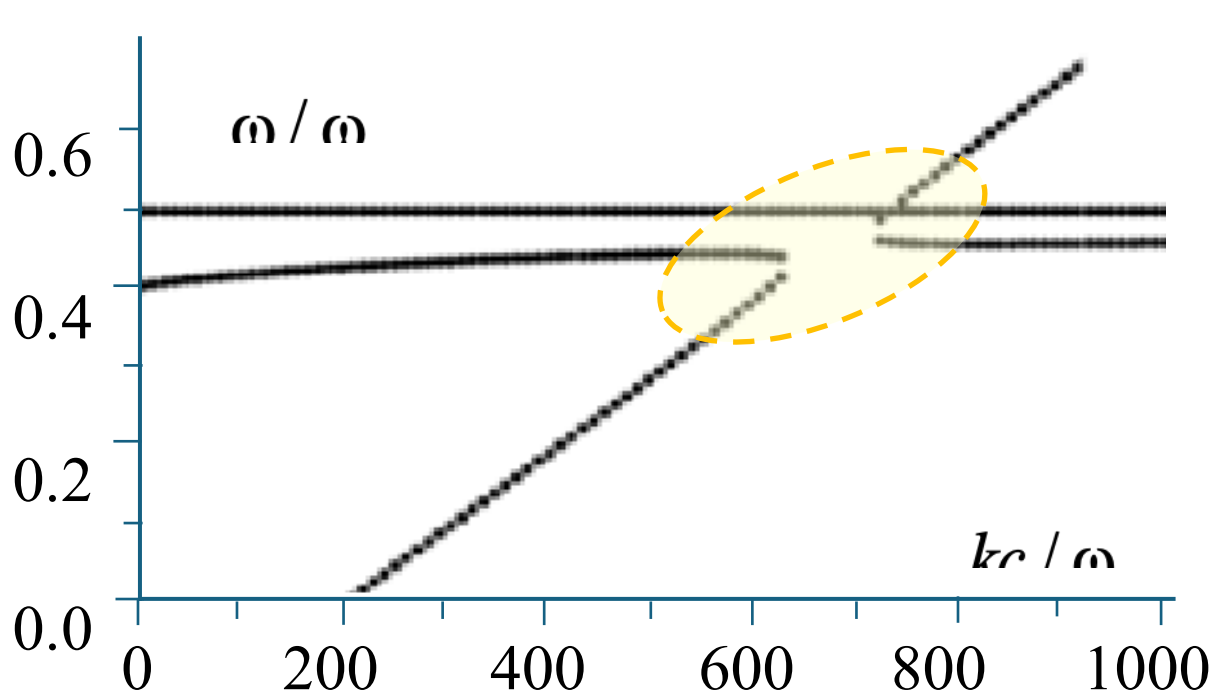

Fig. 2. Dispersion curves of p-polarized surface waves when DC is in a substrate and nanoparticles on the surface of magnetic film are distributed uniformly.

$$\begin{aligned} E_i\left(\mathbf{k}, z, \omega\right) = E_i^{(0)}\left(\mathbf{k}, z, \omega\right) - k_0^2 n G_{ij}^{(33)}\left(\mathbf{k}, z, \omega\right)\Big[ \tilde{\chi}_{jl}^{(0)}\left(\omega\right) E_l\left(\mathbf{k}, z, \omega\right) + \\ + \tilde{\chi}_{jl}^{(+)}\left(\omega\right) E_l\left(\mathbf{k} + \mathbf{w}, z, \omega\right) + \tilde{\chi}_{jl}^{(-)}\left(\omega\right) E_l\left(\mathbf{k} - \mathbf{w}, z, \omega\right) \Big]. \end{aligned} \tag{9}$$

The designations $\tilde{\chi}_{jl}^{(0)}(\omega) = \int\limits_V d\mathbf{r}\chi_{jl}(\mathbf{r},\omega)$, being the averaged susceptibility of a single nanoparticle on the surface, and $\tilde{\chi}_{jl}^{(\pm)}(\omega) = \int\limits_V d\mathbf{r}e^{\pm i\mathbf{wr}}\chi_{jl}(\mathbf{r},\omega)$, being the parameter, which reflects the regularity of the film's coverage by nanoparticles, were used in Eq. (9). Because all nanoparticles are considered identical, i.e. their centers (localized dipole position in the point dipole approximation) lie at the same distance from the surface of the film, below we will omit the argument $z$.

The connection between the Fourier transform of linear response of the system of plasmonic nanoparticles to the external field can be presented via the effective susceptibility [39] as

$$J_l(\mathbf{k},\omega) = -i\omega\varepsilon_0\Xi_{lj}(\mathbf{k},\omega)E_j^{(0)}(\mathbf{k},\omega), \tag{10}$$

where

$$\Xi_{lj}(\mathbf{k},\omega) = \tilde{\chi}_{li}(\omega)\Phi_{ij}^{-1}(\mathbf{k},\mathbf{k}\pm\mathbf{g},\omega) \tag{11}$$

is the effective susceptibility of the system under consideration, with $\Phi_{ij}^{-1}(\mathbf{k},\mathbf{k}\pm\mathbf{g},\omega)$ the so-called local-field factor connecting the local and external field [40]. The local-field factor is calculated by taking into account the processes of a surface plasmonic wave scattering on a Bragg's plane of the 2D photonic crystal. Namely,

$$\begin{aligned}\Phi_{lj}(\mathbf{k},\mathbf{k}\pm\mathbf{w},\omega) &= \mathrm{X}_{jl}^{(P)}(\mathbf{k},\omega) - \\ &-k_0^2n^2G_{il}^{(33)}(\mathbf{k},\omega)\chi_{ik}^{(+)}(\omega)\mathrm{X}_{km}^{(P)}(\mathbf{k}+\mathbf{w},\omega)G_{mp}^{(33)}(\mathbf{k}+\mathbf{w},\omega)\chi_{pj}^{(-)}(\omega) - \\ &-k_0^2n^2G_{il}^{(33)}(\mathbf{k},\omega)\chi_{ik}^{(-)}(\omega)\mathrm{X}_{km}^{(P)}(\mathbf{k}-\mathbf{w},\omega)G_{mp}^{(33)}(\mathbf{k}-\mathbf{w},\omega)\chi_{i'l}^{(+)}(\omega)\ .\end{aligned} \tag{12}$$

From the condition

$$\mathrm{Re}\,\Phi_{ij}(\mathbf{k},\mathbf{k}\pm\mathbf{g},\omega) = 0, \tag{13}$$

one obtains the dispersion relations of a SLR propagating along the surface covered by metal nanoparticles, which form the 2D photon crystal on the surface of a film. The dispersion curves calculated according to Eq.(13) for the lattice constant $L$ = 315 nm are shown in Fig.3. The phase synchronism domains are highlighted. As can be seen, the use of a regular structure led to the emergence of an additional region of phase synchronism ($kc/\omega_p \approx 100 \div 130$) corresponding to wavelengths $\lambda \approx$ 1180 nm. Thus, coatings with a regular arrangement of nanoparticles (2D photonic crystal) can provide an additional control level for the practical application of BLS amplification by a DC. Varying the lattice constant of the photonic crystal, one can achieve the phase synchronism domain for the appropriate wavevector values. An important characteristic of plasmon amplification is the spatial gain increment. It is defined as the imaginary part of the wave vector in the phase synchronism region. Previous evaluations show that the normalized spatial growth increment $\Gamma = \mathrm{Im}(kc/\omega_p)$ can reach a value of 30 at the frequency $\omega/\omega_p \approx 1.1$. Accordingly, pumping energy from a DC source into a surface plasmon resonance is expected to increase the surface-plasmon amplitude by more than an order of magnitude. Indeed, according to Eqs. (4) and (5), one can write $E_i^{(BLS-II)}(\mathbf{R},\bar{\omega}) \propto F\left[(\mathrm{Re}\,B)^2 + (\mathrm{Im}\,B)^2\right]^{-1}$, with $F = \int\limits_{V_{\mathrm{int}}} d\mathbf{R}' G_{ij}^{(33)}(\mathbf{R},\mathbf{R}',\bar{\omega})A_{jl}^{(P)}(\mathbf{R}',\bar{\omega})E_l^{(BLS-I)}(\mathbf{R}',\bar{\omega})$, where $A_{jl}^{(P)}$ is the numerator, and $B$ is the pole part (denominator) of the effective susceptibility of the plasmonic nanosystem. Under the plasmon resonance condition, $\mathrm{Re}\,B = 0$, which implies $E_i^{(BLS-II)}(\mathbf{R},\varpi) \propto F / \left(\operatorname{Im} B\right)^2$. The imaginary part of $B$ is defined by dissipative processes and can be estimated as $\operatorname{Im} B \approx \varpi\Gamma$, where $\Gamma=1/\tau$, and $\tau$ is the relaxation time of the SCR wave. Thus, $E^{BLS-II}(I) \propto F\tau^2 / \varpi^2$. Conversely, when the DC is absent ($I = 0$), $E^{BLS-II}(I=0) \propto F\tau_0^2 / \varpi^2$, where $\tau_0$ is the relaxation time in the absence of DC. A 30-fold increase in the spatial growth increment can be interpreted as an approximate 30-fold increase in the relaxation time. This implies that $E^{BLS-II}(I) \propto F(30\tau_0)^2 / \varpi^2$. Thus, the increase in the amplitude of the BLS field when DC flows along the substrate can reach an upper estimate of $E^{BLS-II}(I) / E^{BLS-II}(I=0) \sim 10^3$. By contrast, plasmon-resonant BLS without an external energy supply ($I = 0$) yields only a several-fold increase. Therefore, introducing this additional interaction can realize true BLS amplification, rather than the standard enhancement typically provided by plasmonic nanoparticles.

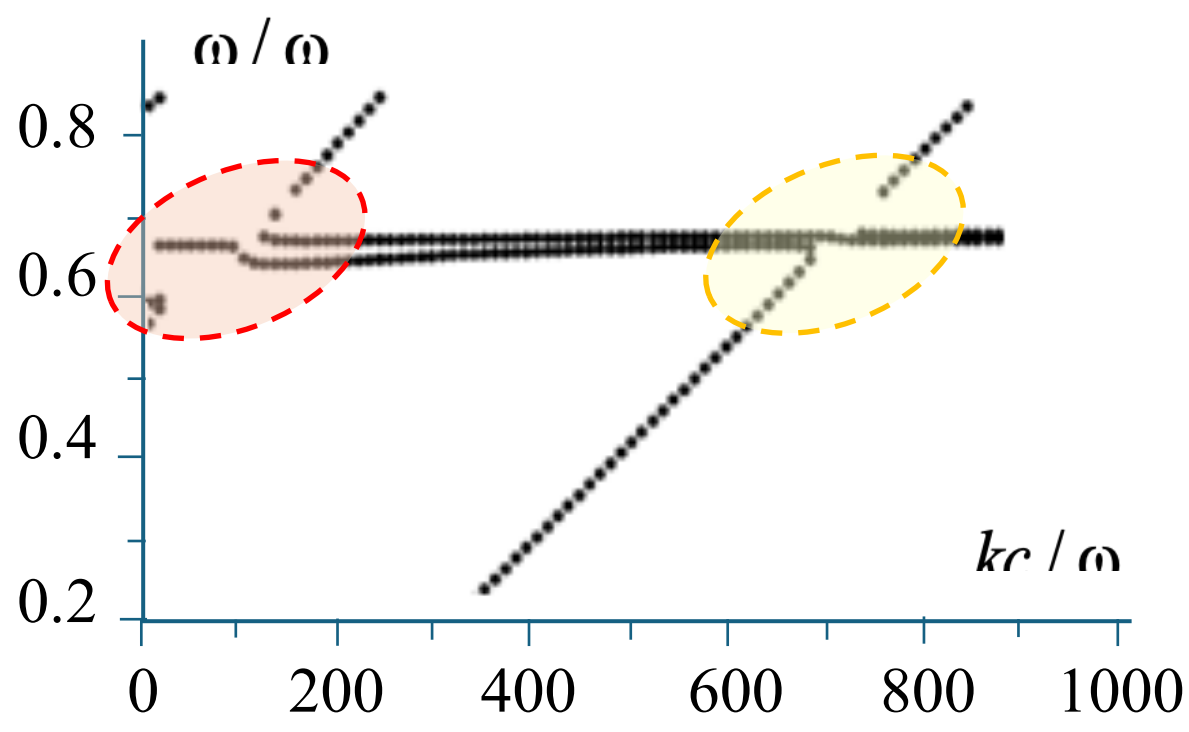

Fig. 3 Dispersion curves of p-polarized surface waves when DC is in a substrate and nanoparticles on the surface of magnetic film form the photonic crystal. SCR wave and DC propagate along OX axis.

As noted in the Introduction, amplification of the surface plasmon via DC-to-plasmon energy transfer can, in formal terms, be viewed as both (i) an increase of the numerator and (ii) a reduction of the imaginary part of the denominator in the effective susceptibility. Either process increases the effective susceptibility and thereby amplifies the BLS signal, in accordance with Eqs. (4) and (5). Although interactions typically raise the imaginary component of the susceptibility, in the present case the added interaction supplies energy from an external source and can extend the common plasmon mode mean free path (i.e., increase its lifetime), thereby reducing the imaginary part of the effective susceptibility.

## IV. AMPLIFICATION OF A SCR BY EXCITATION OF AN AUXILIARY ABSORBING STRUCTURE ON A MAGNETIC FILM

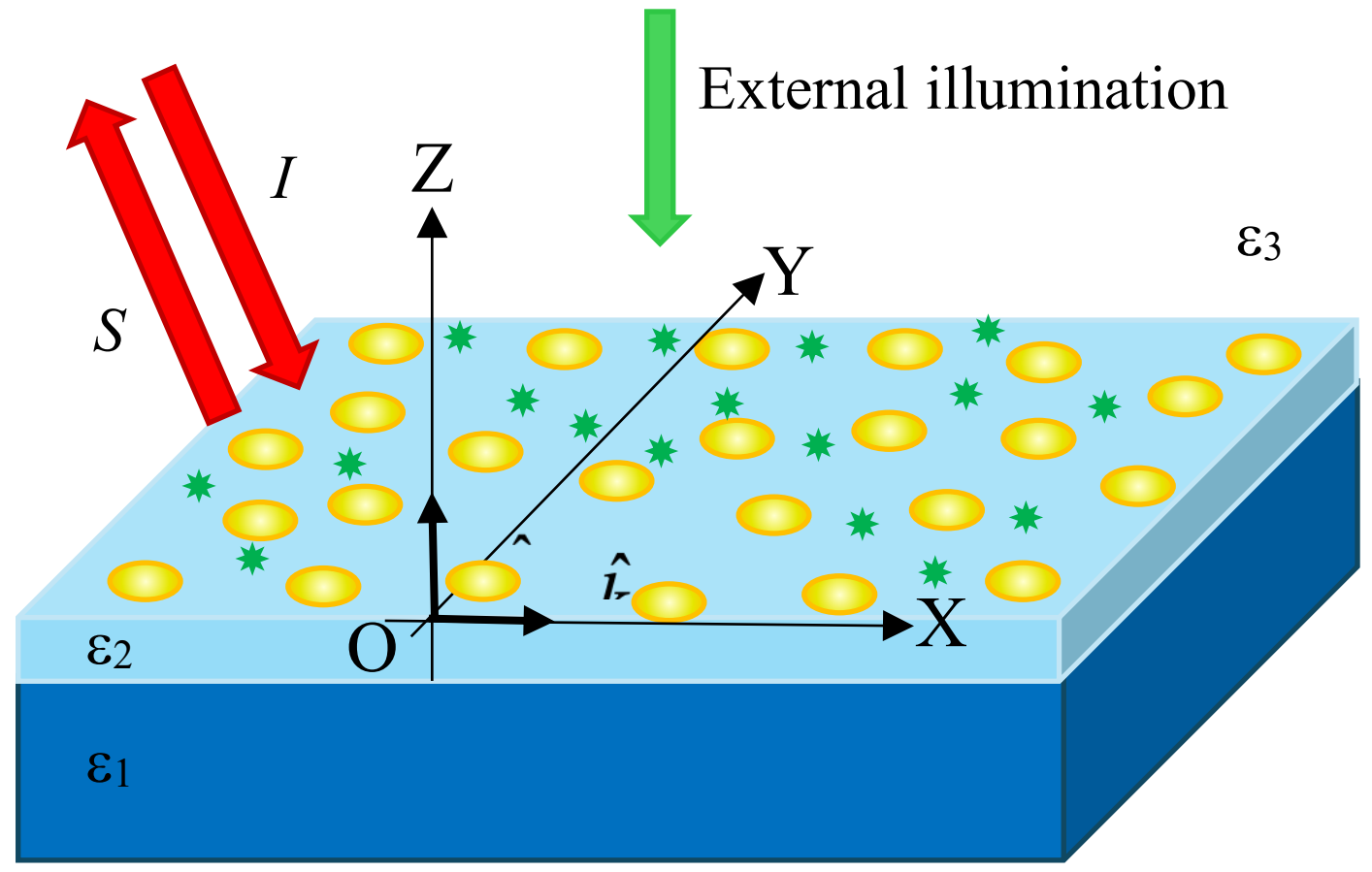

Fig. 4 Setup of the system where the amplification of BLS is by external illumination. Yellow ellipsoids mean the nanoplasmonic particles, green stars mean the absorber particles (organic dye molecules or quantum dots).

In this section, we examine the possibility of the energy transfer from a layer of auxiliary absorbers on a magnetic film (Fig. 4) excited by an additional external light source to SCR. The surface hosts two coupled subsystems: an absorber layer and a plasmonic layer capable of supporting a surface electromagnetic mode. The absorbers – organic dye molecules or semiconductor quantum dots – are co-deposited with the nanoparticles, and are hereafter referred to as the active layer.

The absorbers must efficiently capture the incident light and funnel energy into the SCR by stimulating emission from a quasistatic level. Realization of effective stimulated emission into the common electromagnetic mode of the system of plasmonic nanoparticles demands a corresponding energetical structure of absorbers, similar to the energetical structure of molecules for laser generation. It can be 3-level structure with metastable intermediate level $E_2$ exhibited in Fig. 5 or 4-level structure, which gives more effective population inversion [41]. Effective energy exchange between absorbers and the SCR is provided by the choice of absorbers with emission frequency from the quasistatic level resonantly coincident with the frequency of the SCR. As such a system of two interacting subsystems allows energy flux in both directions, it is clear that for the effective amplification of the SCR instead of its extinction by resonant absorbers, the population inversion is necessary to make the sum energy flux between subsistems in the direction from absorbers to the SCR. This condition determines the corresponding energy structure of the absorber and such active layer acts as a quantum amplifier [42].

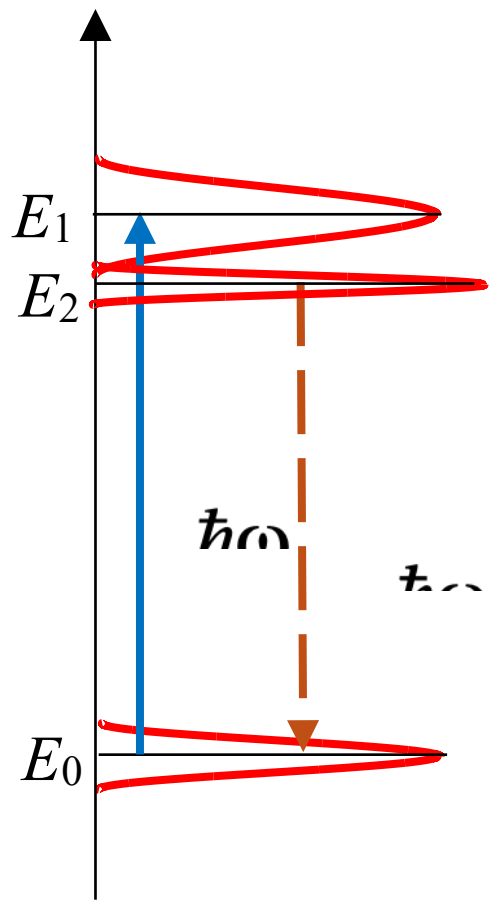

Fig. 5 Sketch of the energy structure of the unit of active layer. $E_0$ is a ground state, $E_1$ and $E_2$ are excited energy levels.

The energy-transfer processes from the external field to the SCR are conveniently treated within a quantum-electrodynamic framework. Following Ref. [43], the dispersion relation of the modified

$$H_{SP} = \sum_{\mathbf{k}} \hbar\omega_{\mathbf{k}} \left( a_{\mathbf{k}}^{+} a_{\mathbf{k}} + \frac{1}{2} \right), \; H_{AL} = \sum_{i} N_i \hbar\omega_i b_i^{+} b_i \; ,$$

$$H' = -\sum_{j,\mathbf{k}} \mathbf{p}_j \mathbf{E}(\mathbf{k}, z) e^{i\mathbf{kr}} \left( a_{\mathbf{k}} + a_{-\mathbf{k}}^{+} \right), \tag{15}$$

where, $a_{\mathbf{k}}^{+}$ and $a_{\mathbf{k}}$ are the SCR creation and annihilation operators, respectively; $\hbar\omega_{\mathbf{k}}$ is the energy of quanta of the SCR, with $\omega_{\mathbf{k}} = \omega(\mathbf{k})$ being the dispersion relation of this mode, given as the solution of Eq.(14). Operators $b_{\mathbf{k}}^{+}$ and $b_{\mathbf{k}}$ are the creation and annihilation operators of the active layer excitations with energy $\omega_i$, $N_i$ is the number of excited units, and $\mathbf{p}_j = \mathbf{p}_{eg}(b_{\mathbf{k}j}^{+} + b_{\mathbf{k}j})$ ($\mathbf{p}_{eg}$ is electric dipole when molecule transit from ground to excited state) is an operator of the dipole momentum at the $j$-th units. Interaction between the SCR and the active layer causes absorption or stimulated amplification of the SCR. Let us consider the energy structure of the unit of the active layer shown in Fig. 5 Let's assume that the lifetime of the excited state $E_1$ of the absorber is remarkably shorter than all other times in the system. so, after the transition from the ground level to level $E_1$, the absorber quickly relaxes either with the probability $p$ to the level $E_2$, where the interaction with the SCR occurs, or to the ground state. Thus we can consider this level permanently empty. The energy of the $E_2 - E_0$ transition is equal to the energy quanta of the SCR. Then it can be assumed that the interaction between the excited unit of the absorber and the surface plasmon occurs according to the Förster mechanism [44]. With $N$ as the surface density of active layer units we can write the rate of change of the population of the units in the level $E_2$

$$\begin{aligned} \frac{\partial}{\partial t} N_2(x,t) = pI_{ext} \left[ N - N_2(x,t) \right]^2 W_{01}(x,t) - N_{SCR}(x,t) \left[ N - N_2(x,t) \right] W_{Pl2}(x,t) - \\ - N_2(x,t) W_{Pl2}(x,t) / S - N_2(x,t) \left[ N - N_2(x,t) \right] W_{20}(x,t) \; , \end{aligned} \tag{16}$$

where $S$ is a unit square used here to keep the same dimensionality of addendums, $W_{01}(x,t)$ is an effective rate of the $0 \rightarrow 1$ transition under the action of the pumping field, which is $I_{ext}$, $W_{Pl2}(x,t)$ is the rate of the exchange between the SCR and elements of the active layer, which, according to the Einstein rule, is equal for the absorption and excitation of the SCR, and $W_{20}(x,t)$ is the rate of the direct relaxation $2 \rightarrow 0$. The rate equation for the intensity of the SCR can be written as follows

$$\begin{aligned} \frac{dI(x,t,\omega)}{dx} = \frac{\partial I(x,t,\omega)}{\partial x} + \frac{1}{v_{gr}(\omega)} \frac{\partial I(x,t,\omega)}{\partial t} = \\ = \left[ \hbar\omega w_{20}(\omega) N_1(x,t) - \hbar\omega w_{02}(\omega) N_0(x,t) - \gamma \right] I(x,t,\omega) \; , \end{aligned} \tag{17}$$

where $\gamma^{-1}$ is the mean free path of the SCR in the film, $v_{gr}(\omega)$ is the group velocity of the SCR. This equation demonstrates that the intensity of the SCR increases due to additional creation of SCR quanta by the units of active layer [first term in the brackets in the right part of Eq.(17)] and decreases due to energy absorption by the units of active layer along with damping in the film (second and third terms in the brackets). Therefore, the amplification of BLS can be observed when

$$\hbar\omega \left( N_2 w_{Pl2}(x,t) - \left( N - N_2(x,t) \right)(x,t) w_{Pl2}(x,t) \right) > 0 \tag{18}$$

confirming that for the SCR amplification we need the population inversion of the level $E_2$ ($N_2 > N/2$). Considering random distribution of both kinds of object, some nonuniform spreading of lines is expected. However, despite noticeable overlapping of both lines is a prerequisite for the effective coupling, exact matching is not necessary, as stimulated emission happens exactly at the frequency of the stimulating excitation.

The increase of the amplitude of SCR per unit propagation distance depends on the pumping intensity of the external source $I_{ext}$ and natural damping $\gamma$. This dependence, for the case where the active layer units were Rhodamine 6G organic dye molecules covering a gold film surface, was calculated in [45]. In particular, the spatial dependence of the number of SCR quanta with the propagation through active layer on the pumping beam intensity $I_{ext}$ changes its behavior from decreasing to increasing. Preliminary estimations for the case with a free path of the SCR $\gamma = 10^4\, cm^{-1}$, Rhodamine 6G molecules concentration $N_0 = 5.5 \cdot 10^{12}\ cm^{-2}$ (which corresponds to submonolayer coverage), the pumping light wavelength $\lambda_p = 532\, nm$ that corresponds to the maximum of the molecules' absorption with a cross-section $\sigma_{abs} = 4.3 \cdot 10^{16}\ cm^2$. The external radiation pumping beam (external energy source) excites the molecules from the ground level '0' to the 1st. The excited molecule nonradiatively goes into the 2nd level with a relaxation time $\tau_{12} \approx 10^{-14} s$ (Fig.5). The induced transitions 2→0 with probability $W_{Pl2}$ provide the process of energy transfer from the excited molecule to the SCR. The surface wave wavelength $\lambda_{SP} = 594\, nm$ that corresponds to the $2 \leftrightarrow 0$ electronic transition in the Rhodamine 6G molecule, and the lifetime of the electron on the 2nd level $\tau_{21} = 2 \cdot 10^{-9} s$ [43] were considered.

## V. DISCUSSION OF FEASIBILITY

To further illustrate the predicted BLS amplification via the DC route, Fig. 6 shows the calculated dependence of both the normalized spatial growth increment and the resulting BLS intensity gain on the electron drift velocity in the semiconductor substrate. As the drift velocity approaches the phase velocity of the SCR, the growth increment rises sharply (Fig. 7a), consistent with the onset of convective instability predicted by Sturrock's criterion. In the practical operating range of $v \approx (0.25–0.4)\times10^{-3}c$, corresponding to current densities of 100–200 A/cm² in InSb, the BLS intensity gain reaches values between one and three orders of magnitude (Fig. 7), in agreement with the upper-bound estimate of $\sim10^3$ derived analytically in Section III. The theoretical framework developed in this work is supported by a growing body of experimental and theoretical evidence from related fields. Several independent lines of research confirm the feasibility of the amplification mechanisms proposed here.

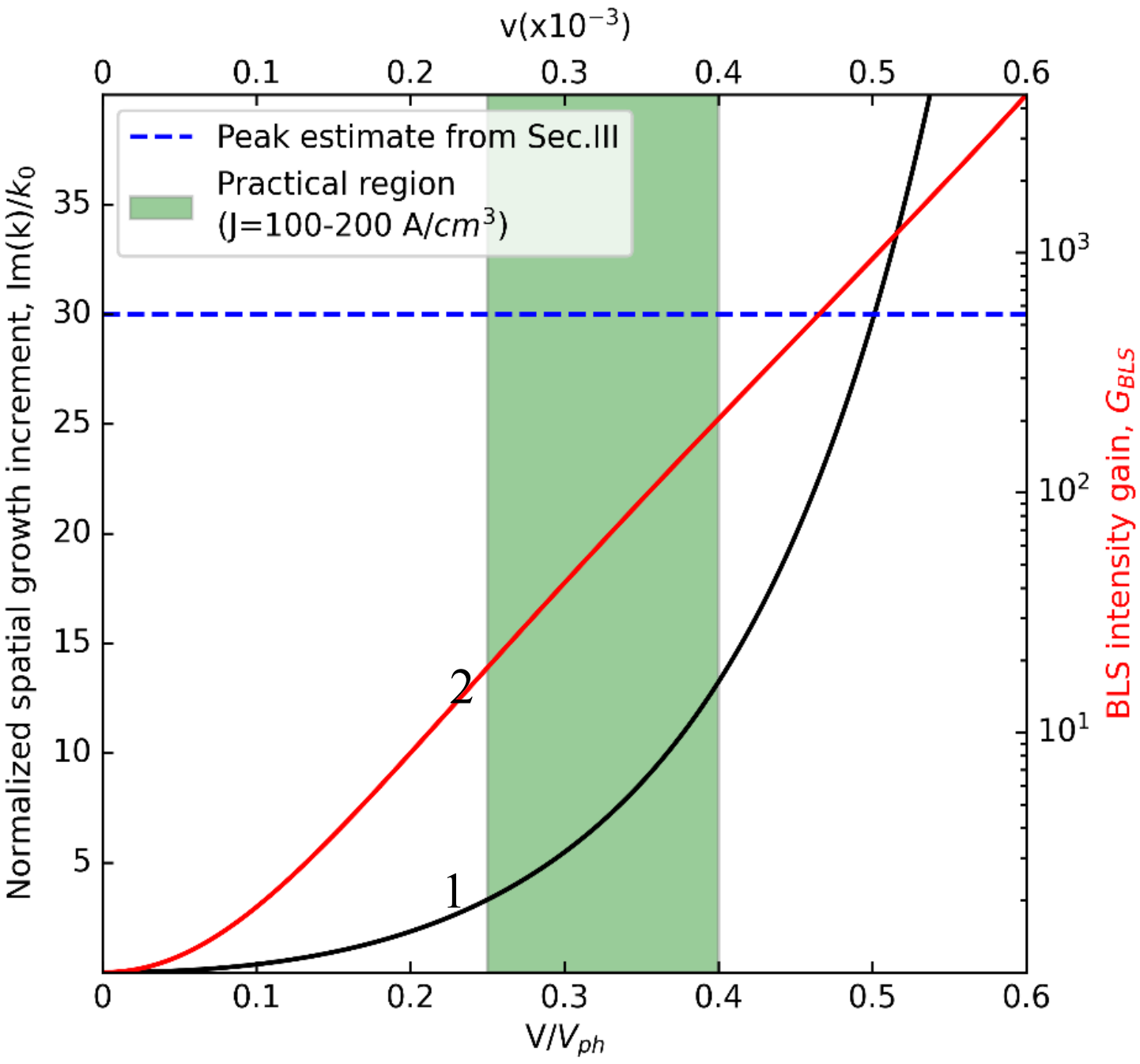

Fig. 6 (1) – Normalized spatial growth increment of the SCR as a function of electron drift velocity. The dashed line indicates the peak estimate of ~30 from Section III. (2) – BLS intensity gain $G_{BLS}$ versus drift velocity (log scale). The shaded region shows the practical range ($j$ = 100–200 A/cm² in InSb).

Note, the concept of amplifying surface electromagnetic modes by direct current in semiconductor substrates has a firm theoretical foundation. The theory of dispersion instabilities associated with surface electromagnetic waves in layered semiconductor media was established in Refs. [34,35]. The theoretical prediction of surface plasmon-polariton amplification was reported in [36]. More recently, the theoretical demonstration that coherent surface plasmon amplification is achievable through the dissipative instability of two-dimensional direct current in graphene-based structures was presented in [48]. Prediction that a DC current in a field-effect transistor channel can lead to plasma wave instability and amplification, was confirmed experimentally at terahertz frequencies in [49]. These results provide strong support for the DC-to-SCR energy transfer mechanism proposed in Section III.

The SCR intensity shown in Fig. 7 was numerically evaluated by solving Eq. (17) for two pumping intensities. When the pumping intensity equals $P_0 = 7.3 \cdot 10^{12}\,\mathrm{erg\,cm^{-1}\,s^{-1}}$, the signal gain is still negative but the total decreasing is smaller than the own damping coefficient γ. This implies that we are already in the amplification regime. When the pumping intensity is increased by a factor of 1.34 to $P_0 = 10^{13}\,\mathrm{erg\,cm^{-1}\,s^{-1}}$, the signal gain is positive and SCR intensity increases (curve 2 in Fig. 7). Thus, an additional pump from an external light source, which excites the auxiliary absorbers at a shifted frequency $\hbar\omega_{ext} = E_1 - E_0$ and provides population inversion, can lead to an increased free path of the SCR. This, in turn, amplifies BLS. Hence, to achieve BLS amplification, it is essential that the area of the surface illuminated by the test radiation be smaller than the area irradiated by the additional external source. This is because the effective interaction region, over whose volume the integration in Eqs. (4) and (5) is performed, is located beneath the test irradiation area.

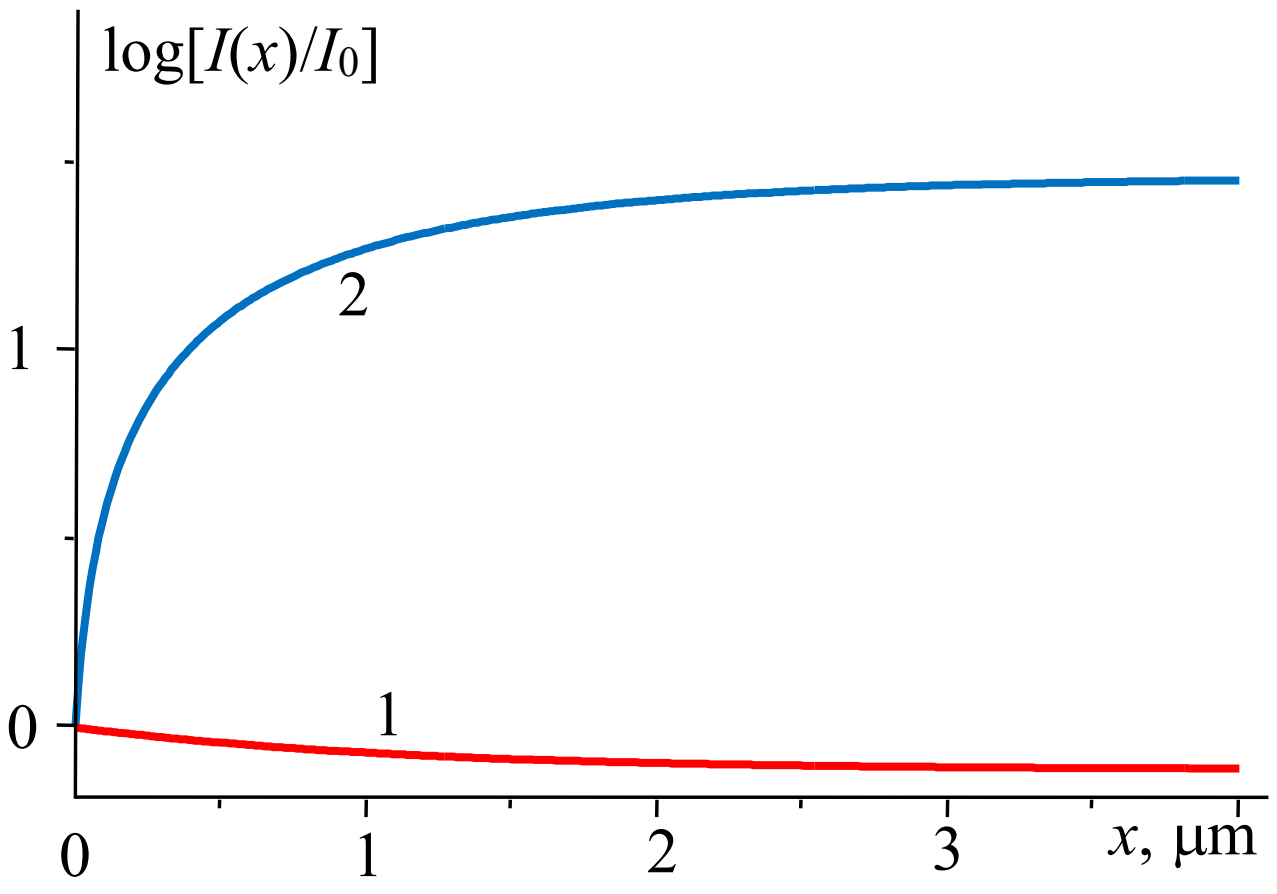

Fig. 7. The dependence of intensity of SP on the length which the wave passes under pumping light action for different intensities of pumping light: Curve 1 corresponds to intensity of pumping light $P_0 = 7.3 \cdot 10^{12}\ \mathrm{erg \cdot sm^{-1} \cdot s^{-1}}$; 2 -

The amplification of surface plasmon polaritons (SPPs) using optical gain media has been extensively studied. The SPASER concept introduced in [50] demonstrated that surface plasmons can be amplified by stimulated emission from a gain medium near metallic nanostructures. In [51], SPP amplifiers and lasers were reviewed, and it was established that organic dye molecules and semiconductor quantum dots can provide sufficient gain to compensate propagation losses. Experimental demonstrations of confined SPP amplifiers using organic gain media confirmed complete loss compensation at moderate pump fluences [52]. These findings validate the amplification mechanism proposed in Section IV.

The integration of plasmonic nanostructures with magnetic systems for enhanced magneto-optical interactions has been demonstrated experimentally. In [11] it was reviewed magneto-plasmonic effects showing that surface plasmon resonances enhance the magneto-optical response by several orders of magnitude. The authors of [53] demonstrated plasmon-mediated magneto-optical transparency. Ref [54] demonstrated nanorod surface plasmon enhancement of laser-induced ultrafast demagnetization exceeding 50%.

Finally, the achievement true amplification of spin waves in magnonic nanowaveguides using spin-orbit torque was reported in [55]. Particularly, amplification demonstrated an exponential increase in spin-wave amplitude up to 5 times. While this mechanism differs from the plasmonic approach proposed here, it establishes that active amplification of spin waves is experimentally achievable. Taken together, these studies confirm that the individual physical mechanisms underlying our proposed BLS amplification scheme are well established. The present work provides a unified theoretical framework combining these mechanisms specifically for BLS amplification in magnonic systems.

## VI. CONCLUSION

Enhancement of the Brillouin light scattering (BLS) signal generated by photon–magnon scattering in a backscattering geometry can be achieved using a nanoplasmonic system on the film surface. Building on our previous works, Refs. [12, 13] demonstrated BLS enhancement when a surface plasmon is excited, and Refs. [34–36, 43] discussed amplification of the surface plasmon via energy transfer from an external source. The present contribution is conceptual and provides a basis for further investigations. We examine two amplification routes. First, energy can be transferred from a direct current (DC) flowing through a semiconductor substrate. Second, additional illumination of an active layer of semiconductor quantum dots or organic dye molecules, co-deposited with the nanoplasmonic system, can transfer energy to the surface plasmon and thereby amplify BLS. The optimal choice is dictated by practical considerations. The DC-driven scheme poses materials challenges. Yttrium iron garnet (YIG) is typically grown on gadolinium gallium garnet (GGG), and its deposition on indium antimonide (InSb) is uncertain. Nonetheless, we expect it may be feasible despite an approximately twofold lattice mismatch. The motivation for InSb is its exceptionally high electron mobility (~70,000 $cm^2 V^{-1} s^{-1}$), enabling electron velocities approaching $10^{-3}c$ under moderate electric fields. By contrast, permalloy thin films are known to be grown on semiconductor substrates [46], suggesting that permalloy on InSb could be a practical option for the DC route. Alternatively, BLS amplification via external illumination of an active layer on the magnetic film surface may be more practical in many experimental scenarios. In all cases, amplification parameters – such as the growth increment of the surface collective mode and the resulting gain – must be evaluated for specific material systems. Detailed calculations for such sytems and further experimental validation will be the subject of future work.

Let's imagine scenarios for such experiments. In the case of BLS amplification by a DC current flowing along the substrate, one should choose a substrate material with high mobility, such as InSb. We set the drift velocity of electrons to $0.3\cdot 10^{-3}c$ (where c is the velocity of light). For moderately doped semiconductor InSb (e.g., $n \approx 10^{14} cm^{-3}$), an applied electric field of approximately 125 V/cm yields a current density of 160 $A/cm^2$, which provides the specified electron drift velocity. Notably, such electric currents induce a magnetic field of about 50 Gauss at the surface of the substrate (inside the magnetic film), which is an acceptable value for BLS experiments.

When BLS amplification is provided by an external radiation source with a submonolayer coverage of active particles, organic dye molecules can be used as active elements. While we previously discussed the Rhodamine 6G organic dye, the organic dye Fluorescein may occasionally be more convenient. Indeed, the excitation maximum of these molecules is at 494 nm, and the emission maximum lies in the range of 515–525 nm. This is particularly suitable for studies of spin waves in Permalloy when the test radiation is in the green part of the spectrum.

Additionally, we want to note that as all proposed amplification mechanisms are connected with the appearance of SCR, estimations made in [47] demonstrated visibility of such a mode since random surface coverage 10-15%, so surface conditions are not changed remarkably. This condition even less demanding in the case of an ordered arrangement. The strongest modes in such a case are Rayleigh anomalies, which are far-field coupling existed at the lattice pitch of the order of the wavelength. So, necessary coverage in this case is at least one order of magnitude less.

## REFERENCES

[1] Flebus B., Rezende S. M., Grundler D., Barman A., Recent advances in magnonics, *J. Appl. Phys.*, **133**, 160401. https://doi.org/10.1063/5.0153424 (2023).

[2] Chumak A.V., Serga A.A., Hillebrands B, Magnonic crystals for data processing, *J. of Phys. D: Applied Physics* **50**, 244001. https://doi.org/10.1088/1361-6463/aa6a65 (2017).

[3] Wang, Y. et al., Fast switchable unidirectional forward volume spin-wave emitter. *Phys. Rev. Appl*., **23**, 014066. DOI: https://doi.org/10.1103/PhysRevApplied.23.014066 (2025).

[4] Chumak, A. V. Magnon Spintronics: Fundamentals of magnon-based computing. In Spintronics Handbook: Spin Transport and Magnetism (eds Tsymbal, E. Y. & Zutic, I.) 247–302 (CRC Press, 2019). DOI: 10.1201/9780429423079-6

[5] Noura Zenbaa et al., A universal inverse-design magnonic device, *Nature Electron*. **8**, 106–115. 10.1038/s41928-024-01333-7 (2025).

[6] Dunagin R.E., Serga A.A., Bozhko D.A., Brillouin light scattering spectroscopy of magnon–phonon thermal spectra of an in-plane magnetized YIG film in two-dimensional wavevector space, *J. Appl. Phys*., **137**, 083901. https://doi.org/10.1063/5.0251149 (2025).

[7] Kyung Hunn Han, Jung Gi Kim, and Sukmock Lee, Brillouin light scattering study of the magnetic hysteresis loop in Fe–Ni/Si(100) film with induced magnetic anisotropy, *Solid State Communications*, **129,** 261-265. https://doi.org/10.1016/j.ssc.2003.10.005 (2004).

[8] Hua Xia, Kabos P., Zhang H.Y., Kolodin P.A., Patton C., Brillouin Light Scattering and Magnon Wave Vector Distributions for Microwave-Magnetic-Envelope Solitons in Yttrium-Iron-Garnet Thin Films, *Phys. Rev. Lett*., **81**, 449-452. https://doi.org/10.1103/PhysRevLett.81.449 (1998).

[9] Serga A.A., Chumak A.V., Hillebrands B., YIG magnonics, *J. of Phys. D: Applied Physics* 43(26):264002. DOI: 10.1088/0022-3727/43/26/264002 (2010).

[10] Heinz B., et al., Propagation of Spin-Wave Packets in Individual Nanosized Yttrium Iron Garnet Magnonic Conduits, *Nano Lett*., **20**, 4220–4227. 10.1021/acs.nanolett.0c00657 (2020).

[11] Maksymov I.S., Magneto-Plasmonics and Resonant Interaction of Light with Dynamic Magnetisation in Metallic and All-Magneto-Dielectric Nanostructures, *Nanomaterials* (Basel) . 5(2), 577–613. doi: 10.3390/nano5020577 (2015).

[12] Lozovski V. and Chumak A.V., Plasmon-enhanced Brillouin light scattering spectroscopy for magnetic systems: Theoretical model, *Phys. Rev.* B, **110**, 184419. https://doi.org/10.1103/PhysRevB.110.184419 (2024).

[13] Demidenko Yu., Vasiliev T., Levchenko K.O., Chumak A.V., and Lozovski V., Plasmon-enhanced Brillouin light scattering spectroscopy for magnetic systems. II. Numerical simulations, *Phys. Rev.* B **111**, 014405. https://doi.org/10.1103/PhysRevB.111.014405 (2025).

[14] Katsunari O. Fundamentals of Optical Waveguides. (Academic Press, 2006).

[15] Weeber J.-C., et.al., Near-field observation of surface plasmon-polariton propagation on thin metal stripes, *Phys. Rev.* B., **64**, 045411. https://doi.org/10.1103/PhysRevB.64.045411 (2001).

[16] Masaya Notomi, Manipulating Light by Photonic Crystals, *NTT Technical Rev*., **7**, 14–23. https://doi.org/10.53829/ntr200909rp1 (2009).

[17] Kazuaki S. Optical Properties of Photonic Crystals. (Springer, 2005).

[18] Wojewoda O., et.al., Observing high-k magnons with Mie-resonance-enhanced Brillouin light scattering, *Commun Phys* **6**, 94. https://doi.org/10.1038/s42005-023-01214-z (2023).

[19] Krčma J., et al., Mie-enhanced micro-focused Brillouin light scattering with wavevector resolution, arXiv:2502.03262v2 [cond-mat.mes-hall] 06 Feb 2025

[20] Agrawal A., et. al., Resonant Coupling between Molecular Vibrations and Localized Surface Plasmon Resonance of Faceted Metal Oxide Nanocrystals, Nano Lett. , **17**(4), 2611–2620, https://doi.org/10.1021/acs.nanolett.7b00404 (2017).

[21] Unser S., Bruzas I., He J. and Sagle L., Localized Surface Plasmon Resonance Biosensing: Current Challenges and Approaches, *Sensors*, **15**(7), 15684-15716. https://doi.org/10.3390/s150715684 (2015).

[22] Pitarke J.M., Silkin V.M., Chulkov E.V. and Echenique P.M., Theory of surface plasmons and surface-plasmon polaritons, *Rep. Prog. Phys*. **70**, 1–87. https://doi.org/10.1088/0034-4885/70/1/R01 (2007).

[23] Auguié B. and Barnes W.L., Collective resonances in gold nanoparticle arrays, *Phys. Rev. Lett*. **101**, 143902. https://doi.org/10.1103/PhysRevLett.101.143902 (2008).

[24] Kravets V.G., Kabashin A.V., Barnes W. L., and Grigorenko A. N., Plasmonic surface lattice resonances: A review of properties and applications, *Chem. Rev*. **118**, 5912 - 5951. https://doi.org/10.1021/acs.chemrev.8b00243 (2018).

[25] Hamdad S., Diallo A.T., Chakaroun M., Boudrioua A., The role of Rayleigh anomalies in the coupling process of plasmonic gratings and the control of the emission properties of organic molecules, *Sci. Rep.*, **12**, 3218. https://doi.org/10.1038/s41598-022-07216-1 (2022).

[26] Baryakhtar I.V., Demidenko Yu.V., Kriuchenko S.V., Lozovski V.Z., Electromagnetic waves in the molecular layers adsorbed on the surface of a solid, *Surf. Sci*., **323**, 142-150. https://doi.org/10.1016/0039-6028(94)00642-3 (1995).

[27] Chegel V., Demidenko Yu., Lozovski V., Tsykhonya A., Influence of the shape of the particles covering the metal surface on the dispersion relations of surface plasmons, *Surf. Sci*., **602**, 1540–1546. https://doi.org/10.1016/j.susc.2008.02.015 (2008).

[28] Bortchagovsky E., Plasmonic coupling and how standard ellipsometry can feel surface plasmon, *J. Vac. Sci. Technol*. B 38, 013603. https://doi.org/10.1116/1.5122267 (2020).

[29] Bortchagovsky E., et. al. , Ordered arrays of metal nanostructures on insulator/metal film: dependence of plasmonic properties on lattice orientation, *Nano Express*, **6**, 025011. https://doi.org/10.1088/2632-959X/adde79 (2025).

[30] Fedorchenko A.M., Kotsarenko N.Ia., Absolute and convective instabilities in plasmas and solids, (Nauka, 1981) (in Russian)

[31] Pozhela Yu.K., Plasma and Current Instabilities in Semiconductors, (Pergamon, 1981).

[32] Sturrock P.A., Kinematics of growing waves, *Phys.Rev*., **112**, 1488-1503. DOI: https://doi.org/10.1103/PhysRev.112.1488 (1958).

[33] Akhiezer A.I., Polovin R.V., Criteria for wave growth, *Sov. Phys. Usp*., **14**, 278–285. https://doi.org/10.1070/PU1971v014n03ABEH004700 (1971).

[34] Wallis R.F., Martin B.G. and Quinn J.J., Theory of amplification and instabilities of surface polaritons in semiconductors, *Physica B+C*, **117-118**, 828-830. https://doi.org/10.1016/0378-4363(83)90665-4 (1983).

[35] Martin B.G., Wallis R.F., Theory of dispersion instabilities associated with surface electromagnetic waves in layered semiconductor media, *Phys.Rev.* B, **32**, 3824-3834. https://doi.org/10.1103/PhysRevB.32.3824 (1985).

[36] Lozovskii V., Schrader S., Tsykhonya A., Possibility of surface plasmon-polaritons amplification by direct current in two-interface systems with 2D Bragg structure on the surface, *Opt. Comm*., **282**, 3257–3265. https://doi.org/10.1016/j.optcom.2009.05.032 (2009).

[37] Cochran J.F and Dutcher J.R., Calculation of the Intensity of Light Scattered from Magnons in Thin Films, *J of Magnetism and Magnetic Materials* **73**, 299-310 https://doi.org/10.1016/0304-8853(88)90095-9 (1988).

[38] Bach M.L., Akjouj A. and Dobrzynski L., Response functions in layered dielectric media, *Surface Science Reports* **16**, 95-131. https://doi.org/10.1016/0167-5729(92)90010-9 (1992).

[39] Lozovski V., The Effective Susceptibility Concept in the Electrodynamics of Nano-Systems, *J of Computational and Theoretical Nanoscience*, **7**, 2077–2093. DOI: https://doi.org/10.1166/jctn.2010.1588 (2010).

[40] Keller O., Local fields in the electrodynamics of mesoscopic media, *Phys.Rep*., **268**, 85-262. https://doi.org/10.1016/0370-1573(95)00059-3 (1996).

[41] [https://en.wikipedia.org/wiki/Population_inversion]

[42] Clerk, A. A.; Devoret, M. H.; Girvin, S. M.; Marquardt, Florian; Schoelkopf, R. J., Introduction to Quantum Noise, Measurement and Amplification. Reviews of Modern Physics. 82 (2): 1155–1208. https://doi.org/10.1103/RevModPhys.82.1155 (2010).

[43] Grynko O., Lozovski V., and Tsykhonya A., Amplification of Surface Plasmon-Polariton by Pumping of an Active Layer at the Surface, conference paper of 2015 IEEE 35th International Conference on Electronics and Nanotechnology (ELNANO), 84-88. DOI: 10.1109/ELNANO35751.2015. (2015)

[44] Schneckenburger H., Förster resonance energy transfer–what can we learn and how can we use it? *Methods Appl. Fluoresc*., **8**, 013001. https://doi.org/10.1088/2050-6120/ab56e1. (2020).

[45] Obeed H.H., Tahir K.J., Ridha N.J., Alosfur F.K.M. and Rajaa Madlol, Linear and Nonlinear optical properties of Rhodamine 6G, IOP Conf. Ser.: Mater. Sci. Eng. 928 072024. https://doi.org/10.1088/1757-899X/928/7/072024 (2020).

[46] Wojewoda O., et al., Phase-resolved optical characterization of nanoscale spin waves, Appl. Phys. Lett., **122**, 202405 (2023)

[47] Bortchagovsky E.G. and Demydenko Yu., Internal reflection of the surface of a plasmonic substrate covered by active nanoparticles, Springer Proc. Phys. 210, "Nanooptics, Nanophotonics, Nanostructures, and Their Applications", Fesenko O. and Yatsenko L. eds., Springer, 2018, pp.243-263. https://doi.org/10.1007/978-3-319-91083-3_17

[48] Igor V. Smetanin Igor V., Bouhelier Alexandre and Uskov Alexander V., Coherent surface plasmon amplification through the dissipative instability of 2D direct current, Nanophotonics, 8, 135–143 (2019). https://doi.org/10.1515/nanoph-2018-0090

[49] Dyakonov M. and Shur M., Shallow water analogy for a ballistic field effect transistor: New mechanism of plasma wave generation by dc current, Phys. Rev. Lett., 71, 2465–2468. (1993). DOI: https://doi.org/10.1103/PhysRevLett.71.2465

[50] Bergman D.J. and Stockman M.I., Surface plasmon amplification by stimulated emission of radiation, Phys. Rev. Lett., 90, 027402 (2003). https://doi.org/10.1103/PhysRevLett.90.027402

[51] Berini P. and De Leon I., Surface plasmon–polariton amplifiers and lasers, Nature Photonics, 6, 16–24 (2012). https://doi.org/10.1038/nphoton.2011.285

[52] Kena-Cohen S., Stavrinou P.N., Bradley D.D.C. and Maier S.A., Confined surface plasmon–polariton amplifiers, Nano Lett., 13, 1323–1329 (2013). DOI: 10.1021/nl400134v

[53] Belotelov V.I. et al., Plasmon-mediated magneto-optical transparency, Nature Communications, 4, 2128 (2013). https://doi.org/10.1038/ncomms3128.

[54] Xu Haitian, Hajisalem Ghazal, Steeves Geoffrey M., Gordon Reuven, Choi Byoung-Chul, Nanorod surface plasmon enhancement of laser-induced ultrafast demagnetization, Scientific Reports, 5, 15933 (2015). https://doi.org/10.1038/srep15933.

[55] Merbouche H. et al., True amplification of spin waves in magnonic nano-waveguides, Nature Communications, 15, 1560 (2024). DOI: https://doi.org/10.1038/s41467-024-45783-1.

**Statements and Declarations**

**Author Contributions**

A.C. and V.L designed the study; V.L. and E.B. developed the methods; All the authors analyzed results and interpreted numerical data. All the authors drafted the manuscript.

**Funding Declaration**

A.C. was funded whole or in part by the Austrian Science Fund (FWF) Project No. 10.55776/I6568;
V.L. was funded in part by Grant of Science and Technology Center in Ukraine (STCU), Project Number 9918 "Magnetism in Ukraine Initiative", awarded to the Institute of Magnetism of NAS of Ukraine and MES of Ukraine.

**Acknowledgements**

This research was funded in whole or in part by the Austrian Science Fund (FWF) Project No. 10.55776/I6568, and by Grant of Science and Technology Center in Ukraine (STCU), Project Number 9918 "Magnetism in Ukraine Initiative", awarded to the Institute of Magnetism of NAS of Ukraine and MES of Ukraine. The authors thank Franz Vilsmeier for valuable discussions.

**Competing Interests**

The authors declare no competing interests.

**Data Availability**

This is a conceptual theoretical study; no datasets were generated or analysed. The numerical results can be reproduced from the equations given in the manuscript.